\documentstyle[aps,prb,multicol,epsf]{revtex}
\input psfig

\begin{document}
\wideabs{
\date{Sept. 7th,1998}
\title{Magnetic bound states in the quarter-filled
ladder system $\alpha^\prime$-NaV$_{2}$O$_{5}$}
\author{P. Lemmens, M. Fischer, G. Els, G. G\"untherodt}
\address{2. Physikalisches Institut, RWTH-Aachen, 52056 Aachen, Germany}
\author{A.S. Mishchenko}
\address{RRC Kurchatov Institute,  123182 Moscow, Russia}
\author{M. Weiden, R. Hauptmann, C. Geibel, F. Steglich}
\address{FB Technische Physik, TH Darmstadt,
64289 Darmstadt and MPI CPFS, 01187 Dresden, Germany}
\maketitle
\begin{abstract}

Raman scattering in the quarter-filled spin ladder system
$\alpha^\prime$-NaV$_{2}$O$_{5}$ shows in the dimerized singlet
ground state (T$\leq$T$_{SP}$=35K) an unexpected sequence of three
magnetic bound states. Our results suggest that the recently
proposed mapping onto an effective spin chain for T$>$T$_{SP}$ has
to be given up in favor of the full topology and exchange paths of
a ladder in the dimerized phase for T $<$ T$_{\rm SP}$. As the new
ground state we propose a dynamic superposition of energetically
nearly degenerate dimer configurations on the ladder.

\end{abstract}
\pacs{PACS: Spin-Peierls transition, spin ladder,
$\alpha^\prime$-NaV$_{2}$O$_{5}$, Raman scattering}

}
\narrowtext


Low-dimensional quantum spin systems like spin chains, ladders or
plaquettes received considerable attention from both theoretical
and experimental points of view due to their manifold of
unconventional spin excitation spectra. Of particular interest is
the so-called spin-Peierls transition. The degeneracy of the
ground state of an isotropic one-dimensional (1d) spin-1/2 system
is lifted due to the coupling to the lattice, leading to a
dimerized singlet ground state and the opening of a
singlet-triplet gap.

The magnetic excitation spectrum of a dimerized spin chain
consists of a triplet branch at $\Delta_{01}$, a corresponding
two-particle continuum of triplet excitations starting at
2$\Delta_{01}$, and well defined magnetic bound states
\cite{uhrig,bouz,Affleck}. The latter consist of strongly
interacting triplet excitations with a high energy cutoff at
2$\Delta_{01}$. The existence of a singlet bound state at
$\sqrt{3}\Delta_{01}$ is predicted by a 1d model with frustrated
next-nearest neighbor interaction $J_2= 0.24 J_1$, acting as a
binding potential. A further triplet bound state should appear for
higher values of $J_2$ \cite{bouz}. A study of magnetic bound
states in quantum spin systems therefore gives valuable insight
into the low energy spin excitations which govern the physics of
these systems.

Half-filled spin ladder systems attracted enormous interest
recently due to the surprising changes of the ground state and
excitation spectrum interpolating between one- and
two-dimensional quantum spin systems \cite{dago96}. For
even-leg ladders the doping by holes is found to lead to
superconductivity. On the other hand, for the quarter-filled
ladder system as the corresponding diluted system the ground
state and spin excitation spectrum are not well understood
until now. However it is expected that they show a similar
rich excitation scheme as the half-filled system
\cite{normand}.

The inorganic compound $\alpha^\prime$-NaV$_{2}$O$_{5}$ with
double chains of edge-sharing distorted tetragonal VO$_{5}$
and a spin-Peierls like transition at T$_{SP}$=35K
\cite{isobe,weiden} has been interpreted as a quarter-filled
spin ladder system \cite{smolinski}. Recent X-ray diffraction
data at room temperature are in favor of a centrosymmetric
(D$_{2h}^{13}$) structure with only one type of V site of
average valence +4.5. The spins are therefore attached to a
V-O-V molecular orbital on the rungs
\cite{smolinski,roth,horsch}. The exchange interaction across
the rung is by a factor of five larger compared to that along
the legs \cite{smolinski}. Hence the ladder can be described
by an effective spin chain.

The formation of a singlet ground state in
$\alpha^\prime$-NaV$_{2}$O$_{5}$ below T$_{SP}$ is observed as
a drop in the magnetic susceptibility \cite{isobe,weiden}. The
observation of superlattice peaks in x-ray scattering proves a
lattice dimerization with a doubling of the unit cell along
the a- and b-axis and a quadrupling along the c-axis
\cite{fuji97}. The singlet-triplet gap was estimated using
magnetic susceptibility: $\Delta_{01}$=85~K \cite{weiden},
NMR: 98~K \cite{Ohama}, ESR: 92~K \cite{vasilev} and neutron
scattering: 115~K \cite{fuji97}. Anomalous behavior of the
spin-Peierls like transition is evidenced by drastic
deviations from the weak coupling regime, e.g. a large value
of the reduced gap $2\Delta_{01}/k_BT_{SP}$=4.8-6.6~K
\cite{weiden,fuji97,Ohama,vasilev}. Moreover, hints for
dynamic two-dimensional spin correlations are observed in
neutron scattering \cite{fuji97}.

In this paper we will present the first experimental investigation
of the spin excitation spectrum of dimerized spin ladders in
$\alpha^\prime$-NaV$_{2}$O$_{5}$ yielding evidence for magnetic
bound states. Raman scattering as a powerful experimental
technique allows to probe dimerization-induced new modes and the
relevant exchange paths by applying polarization selection rules.
Three new modes appearing for T$<$ T$_{\rm SP}$ are identified by
the selection rules and the temperature dependence of their
intensity and linewidth as magnetic bound states. The ground
state, however, is assumed to be composed by a dynamic
superposition of energetically degenerate dimer configurations
along the rungs, legs and diagonals of the ladders.

The scattering geometries available for our phonon Raman
experiments on $\alpha\prime$-NaV$_{2}$O$_{5}$ with our crystals,
i.e. with the polarization of the incident and scattered phonons
parallel to the directions (aa), (bb), (cc), and (ab) one expects
15 A$_{1}$ modes and 7 B$_{1}$ modes for the previously assumed
non-centrosymmetric structure (C$_{2v}^7$) \cite{galy66,golu},
whereas experimentally we observed only 8 and 3 modes
\cite{fisch}, respectively, in agreement with the factor group
analysis for the centrosymmetric structure (D$_{2h}^{13}$). In
addition, infrared absorption data \cite{smirnov,dama} did not
show any phonons coinciding with our Raman-active phonon modes.
Therefore the centrosymmetric D$_{2h}^{13}$ structure should be
favored. Hence there are no two inequivalent V sites which would
give rise to the weakly coupled pairs of V$^{4+}$ and V$^{5+}$
chains as proposed by Isobe et al. \cite{isobe}.

In Fig. 1 we compare the Raman scattering intensity with incident
and scattered light polarization parallel to the legs of the
ladders (b-axis) for temperatures above (100~K) and below (T=5~K)
T$_{\rm SP}$. At high temperature several phonons are observed
ranging from 90 to 1000~cm$^{-1}$ assigned as A$_{1g}$ modes. In
contrast to CuGeO$_{3}$ no spinon continuum is observed with
polarization parallel to the chains. This is in agreement with
previous results that a considerable frustration is required to
observe a spinon continuum in light scattering experiments
\cite{mutu}. In $\alpha\prime$-NaV$_{2}$O$_{5}$ the frustration is
supposed to be weak \cite{smolinski,horsch}. A broad phonon at
422~cm$^{-1}$ ($\approx$ J \cite{isobe,weiden}) hardens in energy
unusually strong by 2.5 \% upon lowering temperature below
T=T$_{SP}$. This demonstrates a non-negligible spin-phonon
coupling in this system.

Several new modes are detected in the dimerized phase in
bb-polarization. The frequencies of these additional
excitations are: 67, 107, 134, 151, 165, 202, 246, 294
(shoulder), 396, 527 (shoulder), 652 (remnant from
aa-polarization), and 948~cm$^{-1}$. In addition, for
frequencies below 120 cm$^{-1}$ an overall drop of the
background intensity is observed. This is a typical phenomenon
for the occurrence of an energy gap. The value of this gap can
be determined experimentally to be about 120-125~cm$^{-1}$. If
we attribute this gap to be the onset of the two-particle
continuum of triplet excitations 2$\Delta_{01}$, we obtain
$\Delta_{01}\approx$ 60-62~cm$^{-1}$, in good accordance with
the value determined from susceptibility data \cite{weiden}.

For the discrimination between phonon modes arising because of
the existence of a superstructure and possible magnetic bound
states in Raman spectroscopy some criteria have been developed
in the case of CuGeO$_3$ \cite{lemmens}. In this frustrated
spin chain compound one singlet bound state is observed only
for incident and scattered light parallel to the chain
direction. These polarization selection rules are consistent
with the spin conserving nature of the exchange light
scattering mechanism ($\Delta$s=0) with a Hamiltonian which is
identical to the Heisenberg exchange Hamiltonian. This process
allows no off-diagonal scattering matrix element (in crossed
polarizations) as observed for one magnon excitations in 2d or
3d antiferromagnets. As the singlet bound state responds very
sensitively by defects and thermal fluctuations its scattering
intensity as function of temperature reaches its maximum
intensity at lower temperatures compared with the dimerization
induced phonon modes \cite{lemmens}. On the other hand, phonon
related bound state phenomena observed, e.g., in rare earth
chalcogenides showed different properties concerning their
selection rules and frequencies \cite{vitins}.

The temperature dependence of scattering intensity, frequency
shift and linewidth for various modes observed in the
quarter-filled ladder system $\alpha\prime$-NaV$_{2}$O$_{5}$
leads us to an unambiguous distinction between the three modes
with energies close to 2$\Delta_{01}$ (67, 107 and
134~cm$^{-1}$) and the other dimerization induced modes at
higher energies. Before pointing this out we have to exclude
that the 67-cm$^{-1}$ mode is a singlet-triplet excitation
(one magnon or spin-flip scattering). This would only be
allowed if magnetic light scattering were dominated by
spin-orbit coupling. In light scattering experiments in a
static magnetic field of up to 7~T neither a shift nor a
broadening of this mode was observed. Therefore a one-magnon
scattering process can be excluded and hence spin-orbit
coupling plays a negligible role in the present context.

Fig. 2a shows the intensity of several modes at low
temperatures (T$<$T$_{SP}$) normalized to the adjacent most
intense ones. Clearly one can distinguish between the modes at
67, 107 and 134~cm$^{-1}$ on the one hand, and the modes at,
e.g., 202, 246, and 948~cm$^{-1}$ on the other hand. The
second group follows the behavior expected for a second order
type phase transition, i.e. the intensity increases quite
sharply upon cooling and then saturates towards low
temperatures. The first group of modes obviously increases in
intensity upon cooling much more gradually and shows no
saturation towards lower temperatures. A similar temperature
dependence of the intensity was found for the singlet bound
state at 30~cm$^{-1}$ in CuGeO$_{3}$ \cite{lemmens}.

Fig. 2b gives further support for this interpretation. The change
of linewidth normalized to the strongest change (67-cm$^{-1}$
mode) is shown as function of temperature. While the three modes
at 202, 246, and 948~cm$^{-1}$ do not show any broadening upon
approaching the phase transition temperature from below, a
behavior typical for simple zone boundary folded phonons, the two
modes at 67 and 107~cm$^{-1}$ become remarkably broader. This
effect is also accompanied by a small softening of their
frequency. The mode at 134~cm$^{-1}$ has been omitted in this
analysis due to a strongly changing background. Nevertheless, the
tendency is the same.

Further insight can be gained from the comparison with a formerly
investigated single crystal of $\alpha^\prime$-NaV$_{2}$O$_{5}$ in
Ref. \cite{fisch} with a slightly broader transition width. The
temperature dependence of the intensity of the additional modes at
64, 103, and 130~cm$^{-1}$ shows a similar distinction from the
other dimerization-induced modes. However, their frequencies are
lower by about 4~cm$^{-1}$ and their absolute intensities are
reduced compared to the single crystals used in the present
investigation. The energy of all other modes is independent of
sample quality. As shown in substitution experiments in CuGeO$_3$
\cite{lemmens} magnetic bound states in quantum spin systems are
extremely sensitive to any defect or thermal fluctuation. Hence
from an experimental point of view we have presented reliable
evidence that the three modes at 67, 107 and 134~cm$^{-1}$ are
related to the singlet-triplet gap and therefore indeed magnetic
bound states.

Spin chain models, which neglect magnetic or magnetoelastic
inter-chain coupling successfully explain the excitation spectrum
of CuGeO$_{3}$ with one singlet bound state \cite{uhrig,bouz}.
Affleck \cite{Affleck} proposed a different physical picture which
allows for more than one bound state. Due to a linear confining
potential mediated via interchain coupling magnetic bound states
of soliton-antisoliton pairs arise. If the slope of this confining
potential is not too strong, there would be a chance for more than
one bound state below the two particle continuum. However, this
would imply a more pronounced one-dimensionality in
$\alpha^\prime$-NaV$_2$O$_5$ compared to CuGeO$_3$.

The experimental data in the dimerized phase do not confirm this.
Fig. 3 shows the polarization dependence of the low energy
excitation spectrum. It can be clearly stated that the intensity
of the 67-cm$^{-1}$ mode is not restricted to bb-polarization
parallel to the legs. It appears in the aa- (i.e. perpendicular to
the ladder) and ab-polarizations, as well. Also the 134-cm$^{-1}$
mode and the onset of a continuum at 120~cm$^{-1}$ are observed in
all three polarization configurations. So the one-dimensionality
concerning the polarization selection rules is obviously not
pronounced. On the other hand, the mode at 107~cm$^{-1}$ fulfills
the selection rules expected for a singlet bound state in a
dimerized chain. It only appears in bb-polarization along the
ladder direction. Also its energy, as pointed out above,
corresponds to the energy of the singlet bound state mode in
CuGeO$_{3}$, i.e., $\sqrt{3}$$\Delta_{01}$=
$\sqrt{3}$(60-62~cm$^{-1}$).

This leads us to the following interpretation of our experiments
based on $\alpha^\prime$-NaV$_{2}$O$_{5}$ as a quarter-filled
ladder: It is not surprising that this compound can have a
spin-Peierls transition into a dimerized ground state assuming an
effective spin chain model. To understand the full excitation
spectrum this point of view obviously is not enough. As more bound
states exist than allowed in models based on dimerized chains we
consider the possible exchange paths of the quarter-filled ladder
to construct further possible dimer states. These might occur
along the rung (a-axis), and the ladder legs (b-axis) and its
diagonals. To obtain the observed superlattice structure and
allowing only weak interaction between ladders the possible dimers
are restricted to a square of adjacent sites. Fig. 4 shows a
representation of these possible states. The energies of a- and
b-axis dimers should be nearly degenerate leading to a system of
several competing ground states. Hopping processes between these
states will form a quantum superposition which splits the dimer
energy levels by $\Delta_{el}$ and leads to a lowering of the
lowest eigenstate. This dynamic configuration is allowed as long
as the energy of phonons $\hbar\omega_{ph}$ with a dominant
spin-phonon coupling that induce the dimerization, fulfill
$\hbar\omega_{ph}\gg\Delta_{el}$. This picture is comparable with
the dynamic Jahn-Teller effect or the RVB model.

In this way the breakdown of the selection rules observed in light
scattering as well as the unexpected low energy of the bound state
at 67~cm$^{-1}$ is described qualitatively. A quantitative
discussion will have to explain the exact energies of the singlet
modes as well as of the modes in the triplet channel. Therefore
neutron scattering data are of crucial importance. In recent
experiments of Yosihama \cite{fuji97} a triplet mode with a steep
dispersion along the b*-axis (ladder direction) splits into two
branches along the a*-axis. Its maximum splitting of 34~K
(=23.5~cm$^{-1}$) gives a new low energy scale and may be
understood as a bonding-antibonding band of the dimers on adjacent
plaquettes. This additional scale is much smaller than the charge
transfer gap $\Delta_{CT}$=0.7~eV of the V-O-V orbital, the lowest
excited state in the spin chain models of Refs.
\cite{smolinski,horsch}. As a rough estimate the energy separation
between the first and the second singlet bound state
(40~cm$^{-1}$) observed in Raman scattering experiments should be
comparable with twice the maximum splitting of the triplet
dispersion along the a*-axis. The factor of two results from the
binding of two triplet states to a singlet bound state.

Using light scattering in $\alpha^\prime$-NaV$_{2}$O$_{5}$,
magnetic singlet bound states were identified. These states
consist of triplet excitations that are bound with respect to the
"free" two-particle continuum of states above 2$\Delta_{01}$.
While the mode at 107~cm$^{-1}$ fits both in energy and selection
rules to a singlet bound state in a dimerized spin-1/2 Heisenberg
chain, the other two bound states do not fit to such a simple
picture. Instead, we propose the origin of the singlet modes at 67
and 134~cm$^{-1}$ as due to a superposition of several nearly
degenerate dimer configurations on the quarter-filled ladder. We
therefore give evidence that, in contrast to the homogeneous high
temperature phase, the dimerized phase of
$\alpha^\prime$-NaV$_{2}$O$_{5}$ cannot be understood on the basis
of a dimerized spin chain. Instead, it is necessary to consider
the topology and enlarged exchange degrees of freedom of a ladder.


Acknowledgment: We acknowledge valuable discussions with C.
Gros, G.S. Uhrig, G. Bouzerar, W. Brenig, P.H.M. v.
Loosdrecht, and G. Roth. Financial support by DFG through SFB
341 and SFB 252, BMBF/Fkz 13N6586/8 and INTAS 96-410 is
gratefully acknowledged.

\begin{figure}
\centerline{\psfig{file=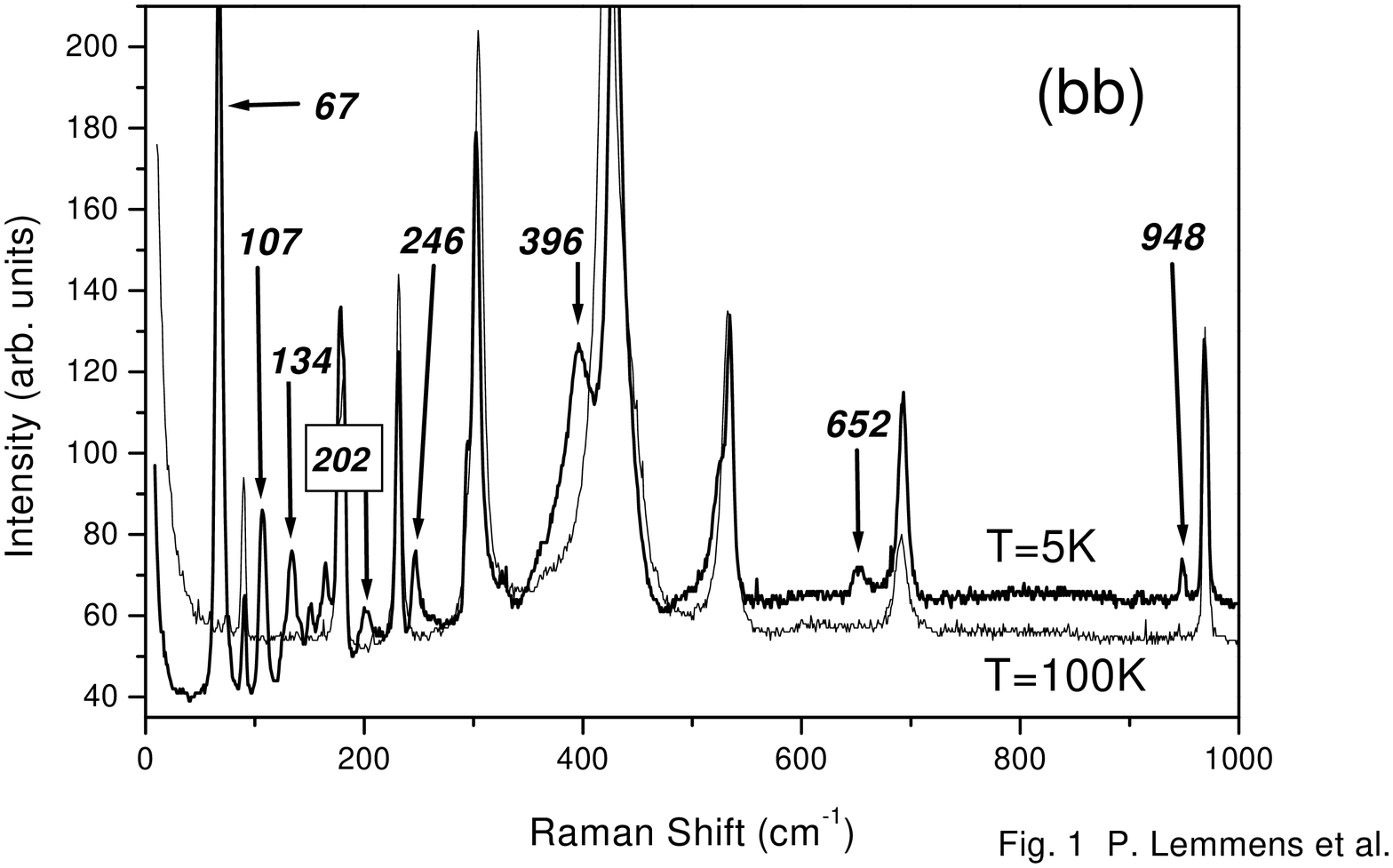,width=10cm}} \caption{Raman
spectra of $\alpha\prime$-NaV$_{2}$O$_{5}$ at 100 (thin line) and
5~K (bold line) with incident and scattered light parallel to the
ladder direction (bb-polarization).} \label{f1}
\end{figure}

\begin{figure}
\centerline{\psfig{file=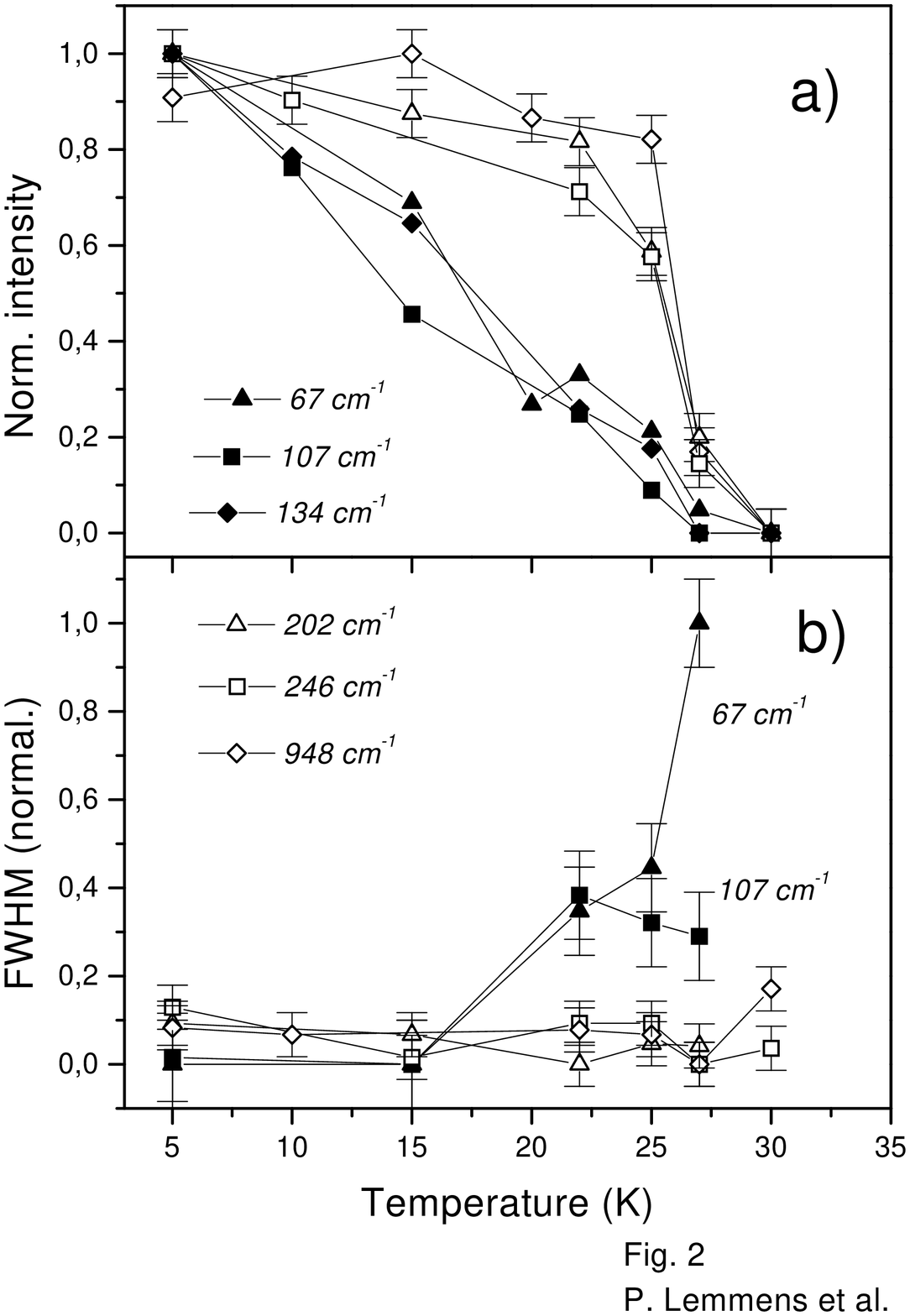,width=8cm}} \caption{a)
Integrated intensity of the additional modes observed below
T$_{\rm SP}$ and b) normalized halfwidth (FWHM) of the additional
modes as function of temperature.} \label{f2}
\end{figure}

\begin{figure}
\centerline{\psfig{file=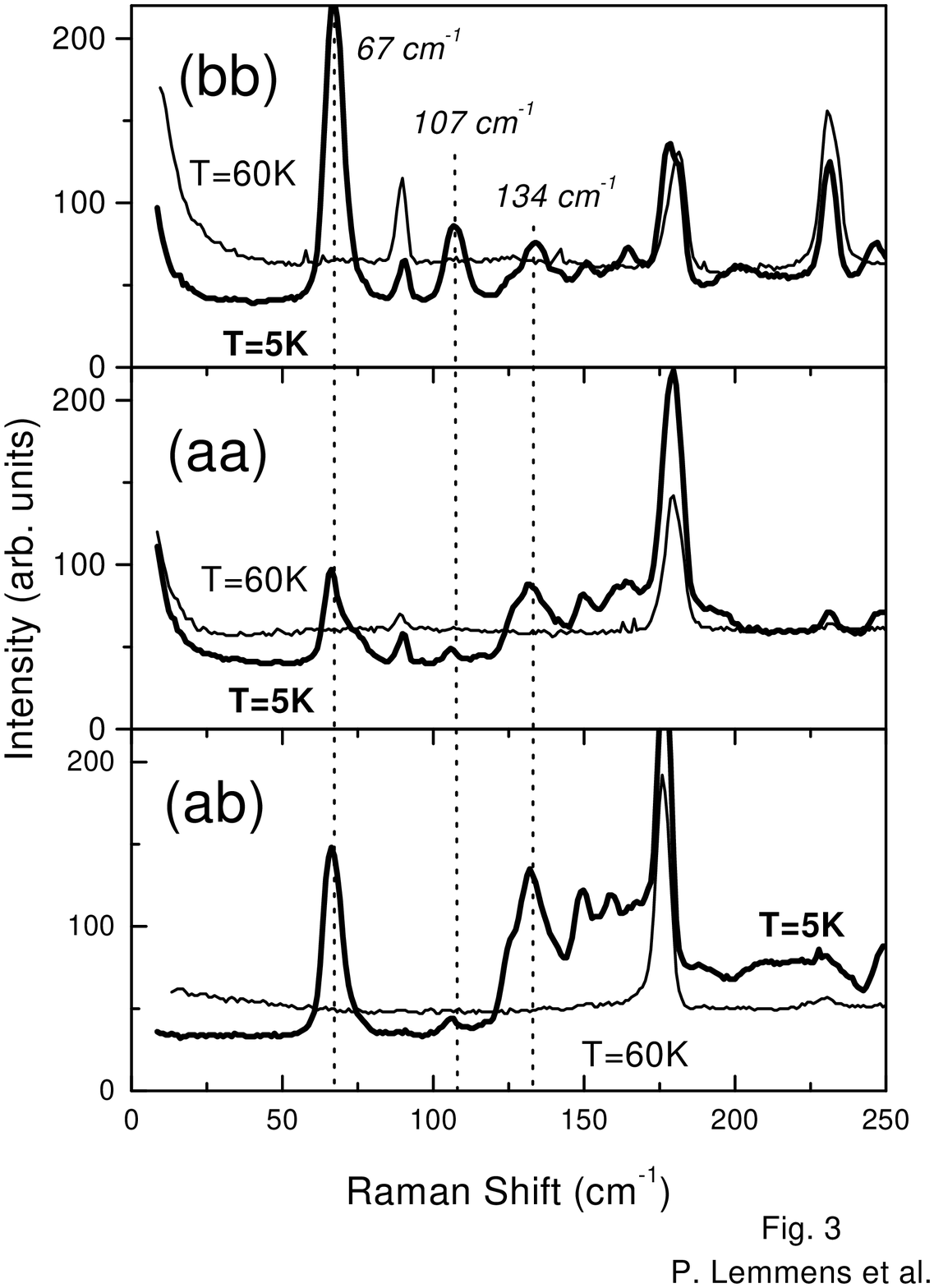,width=8cm}} \caption{Raman
spectra of $\alpha\prime$-NaV$_{2}$O$_{5}$ for three scattering
configurations in the ab-plane of the single crystal.} \label{f3}
\end{figure}

\begin{figure}
\centerline{\psfig{file=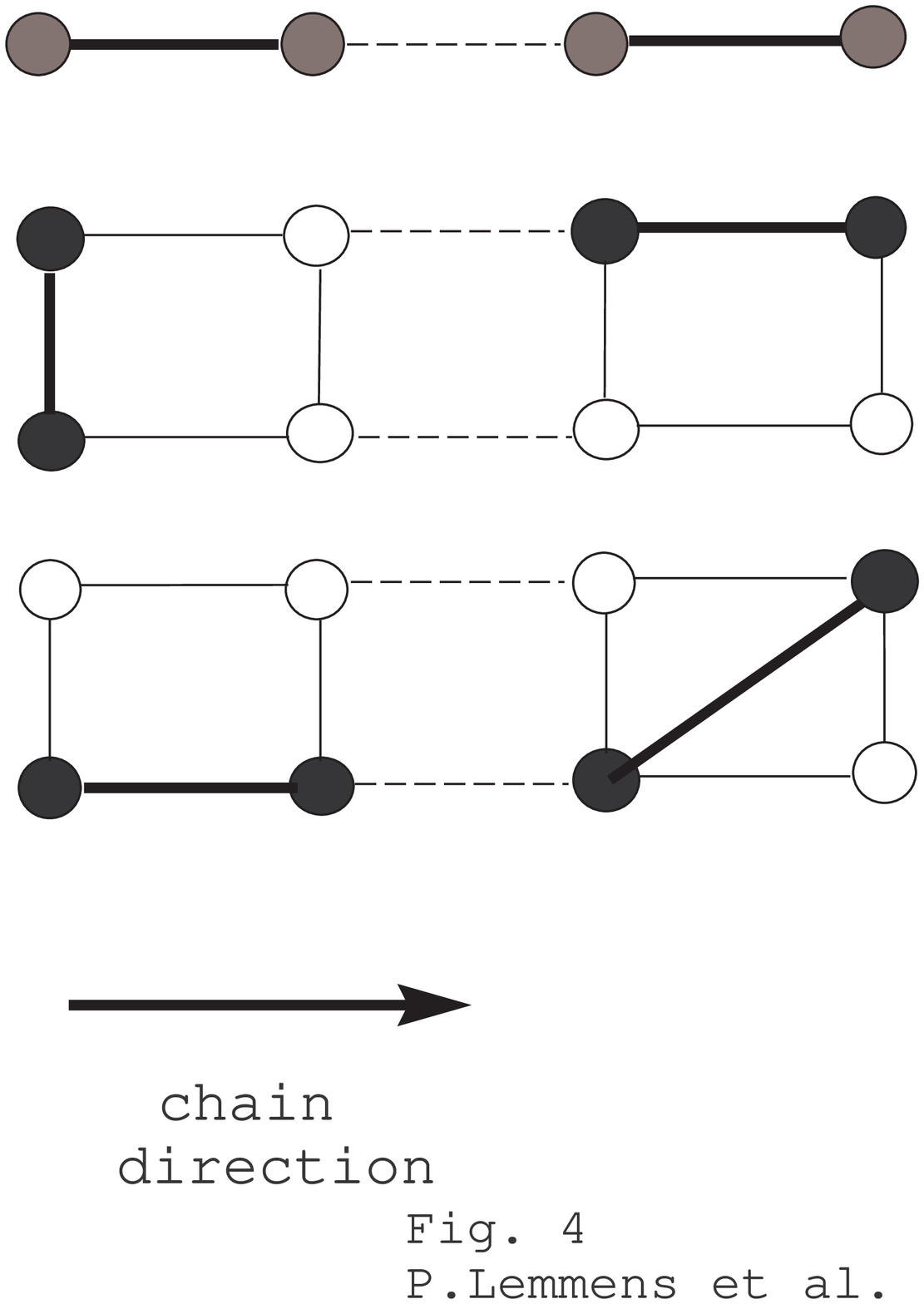,width=6cm}} \caption{Possible
dimer configurations of a dimerized chain (first horizontal row)
and a quarter-filled ladder (second and third row). Possible
microscopic resonating dimer configurations are given. The induced
lattice superstructure is compatible with the observed lattice
doubling in $\alpha\prime$-NaV$_{2}$O$_{5}$.} \label{f4}
\end{figure}

\end{document}